\documentclass[pra,floatfix,preprint,nofootinbib,eqsecnum,superscriptaddress,12pt]{revtex4}
\usepackage{epsfig}
\usepackage{bm}% bold math
\usepackage{amsmath}
\usepackage{enumitem}
\usepackage{color}
\input epsf

\def\be{\begin{equation}}
\def\ee{\end{equation}}

\def\p0{\phi_0}

\def\be{\begin{equation}}
\def\ee{\end{equation}}

\begin{document}
\vspace{1cm}

\title{Computability and Physical Theories}
\author{Robert Geroch}
\affiliation{Enrico Fermi Institute, University of Chicago, Chicago, IL 60637}

\author{James B.~Hartle}
\affiliation{Department of Physics, University of California, Santa Barbara, CA 93106-9530}

%\email{hartle@physics.ucsb.edu}

%\affiliation{Enrico Fermi Institute, University of Chicago, Chicago, IL 60637}
%\affiliation{Department of Physics, University of California, Santa Barbara, CA 93106-9530}

\date{\today}

\begin{abstract}
The familiar theories of physics have the feature that the application of the theory to make predictions in specific circumstances can be done by means of an algorithm. We propose a more precise formulation of this feature --- one based on the issue of whether or not the physically measurable numbers predicted by the theory are computable in the mathematical sense. Applying this formulation to one approach to a quantum theory of gravity, there are found indications that there may exist no such algorithms in this case. Finally, we discuss the issue of whether the existence of an algorithm to implement a theory should be adopted as a criterion for acceptable physical theories. \\

%{\it ``Can it be that there is ... something of use for unraveling the universe to be learned from the philosophy of computer design'?''  \hfill J.A. Wheeler \cite{1} } 

\end{abstract}

\maketitle

%\centerline{``Can it be that there is ... something of use for unraveling the universe to be learned from the philosophy of computer design?'' J.~A. Wheeler \cite{1}
%}

\bibliographystyle{unsrt}

%%%%111111111111111111111111111111111111111111111111111111111111111111111111111111

{\it ``Can it be that there is ... something of use for unraveling the universe to be learned from the philosophy of computer design?''  \hfill J.A. Wheeler \cite{1} } 

\section{Introduction}
\label{intro}

The process of making predictions in physics involves two steps. The first is the formulation of a theory covering a broad class of phenomena; the second is the application of that theory to the specific circumstances of interest. For making predictions of the motion of the planets of our solar system, for example, the first step is the formulation of Newton's theories of mechanics and gravitation; the second the solution of a system of many dozens of ordinary differential equations. These two steps are rather different in character. The first involves the selection, from a wide variety of phenomena, of those one judges to have a common cause, and then the formulation of a mechanism for how that cause is to operate. The test of a good theory is its simplicity and breadth. The second step involves the selection of a specific algorithm by which to extract from the theory its implications. The test of a good algorithm is its convenience.

What underlies our confidence that physics can be structured in this way? Perhaps it is primarily our experience with present-day physical theories. In both classical and quantum physics, the theory culminates in the formulation of differential equations, while the implementation of the theory consists of selecting some algorithm --- of which there are many --- for solving the equations. But not all theories of physics need be of this type. We may someday in physics be confronted with a situation in which the line between the theory and its implementation is not so sharp --- in which the activity of applying the theory is not so different from that of finding the theory in the first place. Specifically, we may reach a point at which there exist no algorithms whatever for applying a theory mechanically to specific circumstances. Indeed, there are suggestions that, with quantum gravity, we may already be at this point.

John Wheeler has stressed on several occasions \cite{1,2} the relation between physical laws, computation, and logic. Here, on the occasion of his 75th birthday, we discuss certain possible formulations of, and attitudes toward, these issues. Section \ref{compu} is a review of mathematics. We briefly summarize the notion of solvability, by computer, of a mathematical problem, and the closely related notion of a computable number. Section \ref{meas}  involves physics. We propose, in parallel with the notion of a computable number in mathematics, that of a measurable number in a physical theory. The question of whether there exists an algorithm for implementing a theory may then be formulated more precisely as the question of whether the measurable numbers of the theory are computable. We argue that the measurable numbers are in fact computable in the familiar theories of physics, but there is no reason why this need be the case in order that a theory have predictive power. Indeed, in some recent formulations of quantum gravity as a sum over histories, there are candidates for numbers that are measurable but not computable. In Sec. 4, we speculate as to the consequences for physics should there be accepted a theory having measurable numbers that are not computable. We suggest that such a theory should be no more unsettling to physics than has the existence of well-posed problems unsolvable by any algorithm been to mathematics.

But, in the spirit of many of John Wheeler's papers, we aim more to raise issues and stimulate discussion than to state conclusions.

%%%%%%222222222222222222222222222222222222222222222222222222222222222222222222222

\section{Computability}
\label{compu}

Computability theory deals with the issue of whether certain well-posed mathematical problems can be solved by means of a digital computer. The key result is a theorem to the effect that one particular problem cannot be solved in this way. From this theorem, one shows unsolvability of a variety of other problems, including that of finding rational approximations to certain real numbers. These are called the noncomputable numbers. We briefly summarize this subject below
\footnote{For a more detailed and rigorous treatment of this material see, e.g \cite{3}}.

We begin with a few ground rules. First, we wish to deal only with digital computers, and not with analog, for we are concerned for the moment only with mathematics, and not with physics. We are interested only in what is possible in principle with such computers, and not in such issues as efficiency or speed. Fix some finite character set, e.g., the standard one of 256 characters. By a string, we mean a finite sequence of characters. Thus, while each string is itself finite, there is no overall upper limit to the allowed lengths of strings. The computer is to have access to an infinite number of memory locations, each capable of holding a string. All these locations are to be blank initially, in order to avoid a situation in which "the answer to the problem" is already encoded, from the beginning, in the memory. The language of the computer is to consist of the specification of certain allowed strings, called the instructions in that language. Of course, not every possible string need be an instruction. A program in the language consists of a finite sequence of instructions. We may represent a program by placing its instructions one after another to form a single long string. Again, not every possible string need be a program. Finally, we demand of the language that it include at least the following three instructions: (i) an input instruction, which temporarily interrupts the running of the program while the user reads in, for possible later use in the program, any string, (ii) a print instruction, which causes the computer to display to the user any string produced by the program, and (iii) a halt instruction, which causes the program to cease execution.

The above is an outline of the types of languages we have in mind. In practice, there are numerous specific languages in which computer programs are commonly written, e.g., Fortran, Basic, Pascal, Cobol, and various machine languages. One might think that whether a problem can be solved by a computer program will depend on the language in which that program is to be written. But, at least for these specific languages, this is not the case: A problem that can be solved in any one of these languages can be solved in any other. Consider, for example, Pascal and Basic. One could write a program in Basic that ``imitates Pascal", i.e., that takes a sequence of Pascal instructions, determines for each what would be done in Pascal, and then does precisely that. Any problem solvable in Pascal is thus solvable in Basic, and similarly for other combinations of these standard languages.

Since, in the sense above, the language is irrelevant, let us fix once and for all a language, and call it ``Fortran," with the understanding that this word can be regarded as standing for any of the others. What problems can be ``solved" by a Fortran program? The idea is to specify in some detail what one wants one's program to do, and then to determine whether or not there exists such a program.

Consider, as an example, Fermat's conjecture: that there exist no positive integers $x, y, z$, and $n$, with $n\ge3$, such that  $x^n+ y^n= z^n$. Does there exist a Fortran program that prints ``true" if Fermat's conjecture is true, or ``false" if it is false, and then halts? The answer to the question, stated in this way, is yes. Indeed, consider the program, with just two instructions, that merely prints "true" and halts; and a second program that prints "false" and halts. Since Fermat's conjecture is either true or false, one of these is the desired program. So, there does indeed {\it exist} a program that does what we have asked. Similarly, any well-posed ``yes-or- no" problem is solvable. One might be tempted to reformulate the question so that its answer requires the specification of {\it which} is the desired program. But it turns out to be difficult to do this in a precise way.

Clearly, the more interesting problems are those that are actually a sequence of problems. For example: Does there exist a program having a single input instruction such that, when any integer $n$ is read in, the program prints ``prime" or ``not prime" according to whether that integer is prime or not, and then halts? The answer of course is yes: Most computer users have written such a program. So, in this sense, the problem of determining whether or not a number is prime is solvable. Similarly, there exists a program that continues to print out the successive digits in the decimal expansion of $\pi$, never halting.

As a final example, we return to Fermat's conjecture, but this time with a different question. Does there exist a Fortran program having a single input instruction such that, when any integer $n\ge 3$ is read in, the program prints ``yes" or ``no" according to whether or not there are positive integers $x$, $y$, and $z$ with $x^n+ y^n= z^n$, and then halts? Since Fermat's conjecture is open, we certainly are not in possession of such a program. [It would not do, for example, to have one's program merely try various combinations of $x$, $y$, and $z$, for then, were $n$ such that there are no suitable combinations, the program would never halt.] Indeed, it is even unclear whether or not there exists such a program\footnote{For purposes of this illustration, we ignore the recent result that Fermat's conjecture holds for all but at most a finite member of n. See \cite{4}.}. There does, for example, if Fermat's conjecture is true (namely, the program that always prints ``no" and halts), or if the conjecture is false for all but a finite number of $n$ (namely, the program that compares the $n$ read in with this finite number of candidates). But what if the $n$'s for which the theorem fails are more ``scattered"?

The key theorem in this subject displays an explicit example of a problem that cannot be solved by any program. It is convenient, for this discussion, to introduce the term {\it monoprogram} to mean a program having a single input instruction. The problem is to determine, given any monoprogram and the string we intend to read in for its input, whether that monoprogram will ultimately halt. In more detail, consider a program X, having just two input instructions and operating as follows. Let, for the first input instruction, there be read in the string representing a monoprogram P, and for the second, an arbitrary string S. Then X itself must always halt, having printed ``halts" or ``does not halt" according to whether or not the monoprogram P, with string S read in for {\it its} input, halts or not. [Should the first string read in to X not represent any monoprogram, then X can do anything.] It is instructive to attempt to write, say, for Fortran, such a program X. One first thinks of various simple arrangements by which P, with S read in, might halt, and so writes X to look for those arrangements. But one continually discovers ever more complicated ways by which P might halt, and so the program X becomes more and more complicated. Eventually, with no end in sight, one gives up in frustration.

There exists no such program X. The proof is simple, but subtle. Suppose, for contradiction, that there were such an X. Construct from it the following monoprogram P. When string S is read in for its input, P merely runs program X, reading in string S for {\it both} of X's inputs. Then P waits patiently for X to finish. If X would have printed "halts," then P goes into an infinite loop, never halting; while if X would have printed "does not halt," then P halts. Thus, given program X one could, by merely changing a few of its instructions as indicated above, write this program P. Now run program P, but, for its input string, read in the string representing the program P itself. Will P so run halt or not halt? Either answer gives a contradiction. Suppose, for example, that P, with the string representing P read in, halts. This means, from the way P was constructed, that X, with the string representing P read in for both its inputs, will print ``does not halt." But this in turn means, from the way X is supposed to operate, that
P, with the string representing P read in, does not halt --- a contradiction. Similarly for the alternative supposition. This contradiction establishes in turn that the desired program X does not exist. This result is called the {\it halting theorem}.

Note that the halting theorem asserts only that there is no general algorithm (i.e., no program X) for deciding whether or not programs halt. There need not be any specific program such that the issue of whether or not {\it that} program halts is undecidable. Indeed, in the formal sense there cannot be, for whether a specific program halts is a ``yes-or-no" question. Imagine that one's job consists of receiving from one's employer a monoprogram and a string, and then trying to decide whether or not that program, with that string read in, halts. The first few cases turn out to be quite easy. Then one day one receives a more difficult case; but, after a month's work, devises a most clever solution. This new method works for the next few cases. But eventually one will receive a case requiring a new, still more clever, solution. The halting theorem states in essence that this
job will never become routine ---  that new and ever more clever solutions will always be required.

It is of interest to see the role played in the halting theorem by the language being used. Clearly, the particular language ---provided only that it has the three basic instructions --- is irrelevant. But it is crucial that the program X be written in the {\it same} language as are the programs P that X tests. Consider, for example, the simplest possible language---that having just the three instructions ``input," ``print," and ``halt." Then the halting theorem in this case asserts that no program X written in this simple language can solve the halting problem for programs written in this same language. One sees directly that this is true, for programs in this simple language cannot ``branch." Suppose, however, that one allows the program X to be written in Fortran. There does exist a Fortran program X that will solve the halting problem for programs P written in our simple language. [Program X merely checks whether or not program P includes a ``halt" instruction.] But, by the halting theorem, no Fortran X will solve the halting problem for Fortran P's. Next, imagine a language that is essentially Fortran, but having an instruction of the form ``Do, for I = 1 to $\infty$," with appropriate modifications to incorporate this into the rest of the language. One can write a program X in this ``super-Fortran" that solves the halting problem for ordinary Fortran programs P. [Program X basically asks whether P has halted after I steps, for ``$I = 1$ to   $\infty$."] But again no such X will solve the halting problem for programs P also written in ``super-Fortran."

While one can imagine the action of ``super-Fortran" language, it is much more difficult to see how one might build a physical computer that
actually runs programs in this language. To execute ``Do, for I = 1 to $\infty$" in the traditional way would require performing an infinite number of steps in finite time. Indeed, to go one step further, it is difficult to see how any language that could actually be run on a physical computer could do more than Fortran can do. The idea that there is no such language is called Church's thesis\footnote{For further discussion of the physical limitations on computing see, e.g  \cite{5}.}. Thus, under Church's thesis, there is a universal notion of an ``algorithm" --- of ``the most that can be done on a physical computer" --- and this notion is correctly captured by standard computer languages. The role of the thesis, then, is to give technical results of the form "... cannot be solved by a Fortran program" the broader meaning "... cannot be solved by any algorithm."

Since we shall shortly be concerned with physics, it is convenient to reexpress the above in terms of the objects with which physics deals: numbers.

A real number $w$ is said to be {\it computable} if there exists a (Fortran) program with a single input instruction such that, when any positive rational number $\epsilon$  is read in, the program prints some rational number r with $|w- r | < \epsilon$ and then halts. Thus, a number is computable if there is an algorithm that generates rational approximations to the number.
[Rationals are used in the definition because they can be ``finitely displayed."] So, for example, the number $\pi$ is computable. Indeed, it is easy to write out an appropriate program in detail, using, e.g., a power-series expansion for $\pi$. Similarly, $\pi^e \sin \sqrt{2}$ ---and, in fact, any other number one can think of off-hand --- is computable.

That there must exist non-computable numbers is easily seen, for the cardinality of the set of Fortran programs (that of the integers) is smaller than the cardinality of the set of real numbers. In fact, one can give an ``explicit example" of such a number. Fix once and for all an ordering (first, second, etc.) for pairs consisting of a monoprogram and a string. [One might, for example, first fix an ordering for the (finite) character set. Then apply dictionary ordering to those pairs for which the program and string combined have just one character, then to those having just two, etc.] Now consider the following number:
\be
\label{e1}
K= \sum_{n=1}^\infty \alpha_n 3^{-n}
\ee
where $\alpha_n$  is one if the $n$th monoprogram with string (in the ordering above) halts, and zero if it does not. This sum, since it necessarily converges, certainly defines a number---in fact, one between 0 and 1. We now claim that this number $K$ is not computable. Indeed, suppose that there did exist a program P to compute it, as above. Then we would immediately have a solution to the halting problem: If you wish to determine whether the $n$th monoprogram with string halts, merely read in to P the value $\epsilon= 3^ {-n-1}$, and, from the rational approximation to $K$ then printed by P, determine whether or not the appropriate term appears in the sum \eqref{e1}. It thus follows, in light of the halting theorem, that K is not computable. More generally, this strategy can always be applied to make the transition between ``whether there exists a program to answer a sequence of yes-or-no questions" and ``whether a certain number is computable."

It is curious that, while the number $K$ above is not computable as we have defined that term, it is ``determinable" in a somewhat weaker sense. We claim: There exists a program that successively prints out an increasing sequence of rational numbers, never halting, such that this sequence converges to $K$. Construct such a program as follows. Fix an integer, say 10. Let our program write out the first ten monoprograrns and strings, run each of these programs for just ten steps, and then evaluate the sum of the first ten terms in (1), letting $\alpha_n$  be 1 if the corresponding program has halted in ten steps, zero otherwise. This is the first rational approximation. The successive approximations are obtained by increasing the integer, say, to 100, then 1,000, etc., and repeating the process. There clearly results an increasing sequence of rational numbers converging to $K$. Note that this program does less than would be required to show K computable, for we do not know at any stage how close the ``rational approximation" is to $K$. If, for example, the first monoprogram required 1017 steps before halting, then our sequence of rationals would "jump upward" by about 1/3 very far along in the sequence. Note also that there can exist no program that prints out a {\it decreasing} sequence of rationals converging to $K,$ for the existence of programs yielding both decreasing and increasing sequences would imply that $K$ itself is computable.

We emphasize that computability is an attribute of the {\it number itself}, and not how that number is presented. Thus, for example, the number $K$ given by  Equation \eqref{e1}  must be irrational, for every rational number is computable. As a further illustration of this point, consider the number
\be
\label{e2}
M= \sum_{n=1} ^\infty 3^{-n}/L(n)
\ee
where $L(n)$ is the number of steps taken by the nth monoprogram and string before it halts (or ``infinite," i.e., omit the corresponding term in \eqref{e2}, if the program does not halt). Is the number $M$ computable? One might think of writing, along the following lines, a program to provide rational approximations to $M$. The program would first determine how many terms in the sum \eqref{e2} would be required to approximate $M$ to within the given allowed error $\epsilon$, would then determine the corresponding $L(n)$, and would finally take the sum of those terms. But this strategy must fail, for, by the halting theorem, there is no program that will determine the $L(n)$. Yet, despite this failure, the number M is actually computable! A program that shows this is the following. In order to approximate M to within error, say,
= 1/100, it suffices to deal only with the first ten terms in the sum \eqref{e2} and, even for these, only either to determine $L(n)$ or else ensure that it exceeds 1,000. So, given $\epsilon = 1/100$, our program merely runs the first ten monoprograms with strings for 1,000 steps each, letting $L(n)$  be infinite for any monoprogram that has not by then halted. Similarly for other $\epsilon$. Thus, a number may appear to be non-computable by virtue of the way it is presented, and yet turn out to be computable by appeal to some more sophisticated algorithm.

\section{Measurability}
\label{meas}

Imagine a physical theory whose prediction, for the result of some particular experiment, is the noncomputable number $K$ of Eq.\eqref{e1}. The theory has thus made a perfectly definite prediction, for $K$ is merely some specific number. Imagine next that we wish to test this prediction. We do so in the usual way: Select some desired accuracy $\epsilon$ carry out the experiment to within that accuracy, and then check to see whether the number that results from the experiment is within $\epsilon$ of K. But, in actually carrying out this test, we would be confronted by the following difficulty: There exists no algorithm to determine whether a given number is within $\epsilon$ of $K$ --- for this is the essence of $K$'s being noncomputable. Of course, it is nonetheless possible to test the theory --- and to do so to arbitrary accuracy. The point is only that the test cannot be carried out mechanically: Each new level of accuracy will require new ideas for the testing procedure. We regard this as the prototype of the situation in which a theory has no algorithm for its implementation. In this section, we formulate this issue more precisely. In the following section, we discuss possible consequences for physics should such a theory come to be accepted.

The key step is the identification of the experimentally determinable numbers that emerge from the theory --- what we shall call the measurable numbers of the theory. This having been done, the issue of whether a theory can be mechanically applied can be stated thus: Are all measurable numbers of the theory computable? It turns out that ``measurable number" must be formulated with some care, in particular to ensure that we thus distinguish only the final physical predictions of the theory, not numbers from intermediate calculations. As a consequence, this term must be given a meaning slightly more restrictive than its common one.

Regard number $w$ as measurable if there exists a finite set of instructions for performing an experiment such that a technician, given an abundance of unprepared raw materials and an allowed error $\epsilon$  is able by following those instructions to perform the experiment, yielding ultimately a rational number within $\epsilon$  of $w.$ ``Measurable" is analogous to, although of course much less precise than, ``computable." The technician is analogous to the computer, the instructions to the computer program, the "abundance of unprepared raw materials" to the infinite number of memory locations, initially blank. Indeed, one can think of the measurable numbers as those that are ``computable" using an analog, rather than a digital, computer\footnote{For other connections between digital computers and the natural "analog" world,see \cite{6}.  The thesis raised in these papers is that limitations on the power of digital computers are universally applicable to their analog counterparts and so can be used to place limitations on the behavior of physical systems. We are not directly concerned with this issue here.}. A few examples will illustrate what we have in mind.

The number $\pi$ is measurable. Suitable instructions for this case might direct that a metal disk be turned on a lathe, that measurements of its circumference and diameter be taken, and that their ratio be computed. Of course, the care with which the disk is to be constructed, and with which the measurements are to be taken, would be dictated by the allowed error $\epsilon$.

The number $e^{1/2}$ is measurable. Suitable instructions might direct that a projectile be released in a viscous fluid, that measurements be taken of
the distance $d_1$ the projectile has traveled when it slows to half its initial speed as well as the ultimate distance traveled $d_2$, and that the combination $d_2/(d_2- d_1) $ be computed. Why would not a similar argument show that $e^K$, where K is the number given by \eqref{e1}, is measurable? It is only necessary to replace in the instructions ``half' by ``a fraction $K$." But these modified instructions would not be regarded as acceptable, for, since $K$ is not computable, it would not be possible for the technician to perform the experiment by merely ``following instructions."

The ratio of the energies of two given levels in the helium atom is measurable. Suitable instructions might direct that there be constructed helium atoms, and that the transition energies be measured with appropriate care.

The fine structure constant is measurable. Suitable instructions\footnote{See \cite{7} for this and other methods of measuring $\alpha$.} might direct that there be measured the wavelengths of the photons emerging from positronium decay as well as that of the lowest energy photon that will ionize hydrogen in its ground state, and that the square root of twice the ratio of these wavelengths be computed.

The meaning of ``unprepared raw materials" is illustrated by the following example. Place two metal rods on a table: Do the obvious instructions show the ratio of their lengths to be measurable? We contend not, for two carefully calibrated rods are hardly ``unprepared." Were we to allow such examples, then clearly any number would be measurable. Measurable numbers are to be those that flow from the operation of physical laws, not from mere initial conditions. The analogous requirement, in the computable case, is that all memory locations initially be blank. Otherwise, any number would be computable. The lathe used to turn the disk in the first example above does not constitute ``prepared raw materials," for one could build a lathe from scratch. The Earth-Moon mass ratio would not be regarded as measurable, since the Earth-Moon system is ``prepared." But we would regard as measurable the proton-electron mass ratio.

All numbers in physics are essentially dimensionless, for even for example an elapsed time reported in seconds merely gives the ratio between that elapsed time and the elapsed time that is the standard second. But, for the issue of measurability, the choice of standard is important. If, for example, the standard second is 1/86,400 of the mean solar day, then instructions reporting times in these units would not be acceptable, for the experiment involves ``prepared" raw materials. But this problem does not arise if the standard second is the time of 9, 192, 631,700 vibrations of a certain hyperfine transition in Cs133. It is curious that most such ``unprepared" standards arise via quantum mechanics.

It is intended that ``measurable" be an attribute of the number itself, and not of how that number is presented. Thus, a given number is either measurable or it is not, the instructions merely serving (as does the computer program in the computable case) to show it measurable. Consider, for example, instructions that direct the technician to work with an unstable system, so the number finally reported varies wildly from one run to the next. There is no number that these instructions show to be measurable. This is analogous to a Fortran program that prints rationals that vary wildly from one $\epsilon$  to the next. There is no number that this program shows to be computable.

The notion ``measurable" involves a mix of natural phenomena and the theory by which we describe those phenomena. Imagine that one had access to experiments in the physical world, but lacked any physical theory whatsoever. Then {\it no} number {\it w} could be shown to be measurable, for, to demonstrate experimentally that a given instruction set shows {\it w} measurable would require repeating the experiment an infinite number of
times, for a succession of $\epsilon$'s approaching zero. One could not even demonstrate that a given instruction set shows measurability of any number at all, for it could turn out that, as $\epsilon$  is made smaller, the resulting sequence of experimentally determined rationals simply fails to converge. It is only a theory that can guarantee otherwise. The situation is analogous to that of trying to demonstrate that a given Fortran program shows some number to be computable. There is no general algorithm for deciding this. In particular, it would not do merely to run the program for a few selected values of $\epsilon$.

So, "The number {\it w} is measurable." is basically a statement about the theory within which we are working, and only indirectly --- only to the extent that the theory correctly describes the physical world --- a statement about natural phenomena. Thus, in the example of measurement of $\pi$, the theory is that of classical continuous elastic solids in Euclidean space. Classical continuous elastic solids is required to ensure that arbitrarily accurate measurements can, in principle, be made on the disk (without interference, for example, from quantum mechanics or atomic structure). Euclidean geometry ensures that the ratio obtained will be independent of disk size, and also that $\pi$ is actually the number to which the approximations converge. In the example of $e^{1/2}$, the theory is that of viscous fluids, and in particular its prediction of exponential decay. In the example of the helium atom, the theory is quantum mechanics; in that of the fine-structure constant, quantum electrodynamics.

Every computable number is measurable. This is easy to see: Let the instructions direct that the raw materials be assembled into a computer, and that a certain Fortran program--one specified in the instructions--be run on that computer. That is, every digital computer is at heart an analog computer.

We now ask whether, conversely, every measurable number is computable --- or, in more detail, whether current physical theories are such that their measurable numbers are computable. This question must be asked with care. Consider, for example, the fine-structure constant $\alpha^{-1}$. This number is measurable in quantum electrodynamics, as discussed above. Is $\alpha^{-1}$  computable? The problem is that in quantum electrodynamics the actual value of $\alpha^{-1}$ is not specified by the theory, but rather is left to be determined by experiment. There is no "number," mathematically specified, whose computability can be investigated. There is no meaning to be attached to the statement  "$\alpha^{-1}$ is computable." 
[Of course, were quantum electrodynamics reformulated with $\alpha^{-1}$ specified, say to be 137.036 (exactly!), then the computability of the fine-structure constant would be meaningful.] We are thus led to divide the measurable numbers into two classes: those assigned specific values by the theory, and those not. 
For example, $\pi$, as measured by the disk, and $e^{1/2}$, as measured by the viscous fluid, are clearly assigned values by the theories. The example of the helium atom is more subtle. If the theory is that of a helium atom with infinite mass, point like nucleus and no relativistic effects, then the ratio of energy levels is assigned a specific value by that theory. But a finite mass brings in the electron-nucleon mass ratio, and relativistic effects the fine-structure constant, neither of which is assigned a specific value in conventional theory. Of course, specific values may be assigned in some future theory.

We suggest that the familiar theories of physics are such that all the measurable numbers specified by these theories are in fact computable. The following example, representing a "typical" theory whose mechanism is a differential equation, will support this suggestion. Let the state of the physical system be characterized by a single variable $x$, evolving with time according to the equation\footnote{The right side of \eqref{e3}  could as well be any smooth "explicit" function--or even some not so explicit. We here merely select one as an example in order to avoid the issue of what is the class of allowed functions.}.
\be
\label{e3}
\frac{dx}{dt}=5(x^2- t^2)\sin(xt) .
\ee
Let it be the case that a technician, given any allowed error $\epsilon$, is able to (i) set the system initially in state $x$ within $\epsilon$ of any given rational number, and (ii) wait for time $t$ within $\epsilon$ of any given rational number, and then measure to within $\epsilon$ the final state of the system. Then under these ground rules the measurable numbers of the theory are clearly the following: all numbers of the form $x(t_0)$, where $t_0$ is any computable number and $x(t)$ is the solution of \eqref{e3} with initial condition any computable $x_0$. Are all of these numbers computable? This is a purely mathematical question, involving the character of the solutions of \eqref{e3}. It turns out that all these numbers are indeed computable, as one sees by noting that one could integrate \eqref{e3} numerically on a computer. Thus, at least in this model theory, the measurable numbers are computable. Note that, were the coefficient 5 on the right in \eqref{e3} replaced by a constant $\lambda$ not further specified, then the measurable numbers would not be specified by the theory; and were it replaced by the non-computable number of \eqref{e1}, then the measurable numbers would not be computable. In a more realistic treatment, the issue of what the technician can set and what he can measure would be determined by the interactions in the theory. We note, finally, that a similar argument applies to field theories governed by hyperbolic equations\footnote{For examples of lack of computability in solutions of differential equations, see \cite{8}.} although the technician must for such examples be given considerable power to control initial conditions.

Thus, we argue, the conventional theories of physics do have the property that all measurable numbers specified by these theories are computable. Here, then, is at least one sense in which it is true that these familiar theories can be applied by means of algorithms.

There is a recent example \cite{9}  in which this issue comes very much to the fore. Consider one version of quantum gravity for closed cosmologies, formulated as a sum over histories. An observable in the theory is a functional, $A[\cal G]$, of geometries $\cal G$  on compact 4-manifolds. The expectation value of such an observable in the analog of the ground state is given, formally, by
\be
\label{e4}
\langle A\rangle=\frac{\sum A[\cal G]  \exp{(-I[\cal G])} }{\sum  \exp{(-I[\cal G])} }
\ee
where $I[\cal G]$ is the Euclidean gravitational action. The sums in Eq.\eqref{e4} are over all compact 4-geometries --- including in particular all possible topologies for the 4-manifolds. Now fix a "reasonable" observable $A$. Is the number given by Eq. \eqref{e4} measurable? Is it computable?

As we have seen, even for simple theories the issue of whether a given number is measurable can be subtle. But this theory is anything but simple, for the theory purports to provide the quantum description of the entire Universe. What instructions would allow a technician to "measure" $\langle A\rangle$? Our understanding of the physics of quantum gravity is not yet to the point at which we can answer this question. Nonetheless, one expects that there will be some measurable numbers in the final theory, and such expectation values would seem to be good candidates.

Is $\langle A\rangle$ computable? Unfortunately, the mathematics of quantum gravity has not yet been developed to the point that we have a complete mathematical formulation of what Eq. \eqref{e4} is to mean. But one possible meaning that has been proposed involves use of Regge calculus. Fix an integer $n$. Consider a simplicial 4-manifold with exactly $n$ vertices. That is, take $n$ points as vertices, and specify which pairs of vertices are to be connected by an edge, which triples of adjoining edges are to be connected by a face, which quadruples of adjoining faces are to be connected by a tetrahedron, and which quintuplets of adjoining tetrahedra are to be connected by a 4-simplex --- all such that the resulting space is a topological 4-manifold. Next, specify numerical values for the lengths of all the edges of this simplicial 4-manifold. The result is a simplicial 4-geometry $\cal G$. We can now evaluate $A[\cal G]$ and $ I[\cal G]$, and thus one term in each of the sums in
 Eq. \eqref{e4}. To evaluate the sums themselves, we proceed as follows: first keep the simplicial 4-manifold fixed and integrate over the values of the edge lengths; and then sum the result over all simplicial 4-manifolds with exactly n vertices. There results the $n$-vertex simplicial approximation $\langle A\rangle$, a number we denote $\langle  A\rangle _n$. Finally, take the limit of the $\langle A\rangle _n$ , as n goes to infinity--assuming this limit exists! The result is our candidate for the expectation value $\langle A\rangle$.

This particular meaning for \eqref{e4} has the following curious feature. Imagine an attempt to evaluate the $\langle A\rangle_n$ by computer. One would select and read in some value for the integer $n$. The computer would then consider $n$ vertices, and try all combinations of which vertices are to be connected by edges, which edges by faces, etc. From the resulting list of simplicial complexes, the computer would then eliminate all those that are not 4-manifolds. Then, from the resulting list of simplicial 4-manifolds, the computer would eliminate all duplications---cases in which there is included more than one simplicial representation of a single 4-manifold. The computer would then perform, for each simplicial 4-manifold in this reduced list, the integrations over edge lengths. Finally, the computer would take the sums over the 4-manifolds. The result would be reported as
$\langle A\rangle_n$ The curious feature is that there exists no computer that can carry out this task! The key problem is with the elimination of duplications. It is known that the issue of whether two simplicial 4-manifolds are topologically identical is undecidable. The theorem, in more detail, is the following \cite{10}.  Let simplicial 4-manifolds be specified by giving the number $n$ of vertices, which vertices are to be connected by edges, which edges by faces, etc. This information could be encoded, for example, into a string. Then: There exists no Fortran program that will accept as inputs two such strings, ultimately print "same manifold" or "different manifolds" according to whether the 4-manifolds represented by the strings are topologically the same or different, and then halt. Further, even the elimination of those simplicial complexes that are not manifolds may be problematic. It appears likely that the question of whether a simplicial complex is a manifold is decidable in dimension four, but undecidable in higher dimensions\footnote{For a review of these results  see \cite{10}.}.
 
 All this is not to say that quantum gravity will admit measurable numbers that are not computable. The mere fact that one particular strategy for writing a program to compute $\langle A \rangle_n$ the fails does not mean that the $\langle A\rangle_n$  are noncomputable. We have seen an example in \eqref{e2}. Further, even should the $\langle A\rangle_n$ turn out to be noncomputable, this does not guarantee that their limit, $\langle A \rangle$, is also noncomputable. It is easy to write down examples of sequences of noncomputable numbers having computable limits. [The noncomputable numbers are dense in the reals.] It could also
 turn out that the meaning ultimately assigned to \eqref{e4} is different from that indicated here. For example, one might simply adopt the rule that duplications of 4-manifolds are not to be excluded in the sums; or, alternatively, the rule that the sums are not to be over different 4-manifolds. It could turn out that the numbers ultimately measurable in the theory are quite different from these expectation values. Or, finally, what is ultimately quantum gravity may bear little resemblance to a sum over simplicial histories.
 
Yet quantum gravity does seem to be a serious candidate for a physical theory for whose application there is no algorithm\footnote{For a discussion and other examples of undecidable problems in physics and mathematical physics see \cite{11} and \cite{12}.}.
 
 \section{Discussion}
 \label{discuss}
 
What must be the character of a physical theory in order that it be regarded as satisfactory? Surely, it must include some mechanism for making quantitative predictions --- or producing, under specified experimental arrangements, quantities subject to experimental verification. Otherwise, we would hardly have a ``theory." But this is by no means the only criterion. The familiar theories of physics possess, in addition to predictive power, such features as simplicity, generality, and elegance. Historically, these criteria have not always been incidental, for they have at times played a central role in the search for physical laws. General covariance is an example --- one for which, remarkably, the ``criterion" turned out to be empty!

We here raise the possibility of another type of criterion: that the theory be such that its predictions can be extracted not merely ``in principle, possibly by ever higher levels of skill and sophistication, but also ``in practice, mechanically." The criterion, in the present more precise version, is that all the measurable numbers specified by the theory be computable numbers. At issue is the status of this criterion.

First note that it may be difficult to decide whether or not a given theory satisfies this criterion. One mathematical formulation of the theory may provide no algorithm for implementing the theory, and yet another, equivalent, formulation does. The physical issue of whether a number is measurable, as well as the mathematical issue of whether it is computable, can be a subtle matter. Nevertheless, it should be far more straightforward to decide whether a theory satisfies this criterion than, for example, to decide whether a theory is ``elegant." It is at least more a question of fact than of taste.
 
 In present physical theories, the computability of the measurable numbers is manifest, so this issue is rarely raised. But now the issue may be forced upon us, by quantum gravity. There are indications that this theory may have measurable numbers that are not computable, at least if its formulation as a sum over simplicial histories is a good guide.
What, then, would be the consequences for physics should there be accepted a theory whose measurable numbers are not computable? Fix some experimental arrangement, and attempt to extract from such a theory its prediction for that experiment. To predict to within, say, 10\%, one manipulates the mathematics of the theory for a while, arriving eventually at the predicted number. To predict to within 1\%, it will be necessary to work much harder, and perhaps to invent a novel way to carry out certain mathematical steps. To predict to 0.1\% would require still more new ideas. The process would be similar to that of trying to compute rational approximations to the noncomputable number given by \eqref{e1}. The theory certainly "makes definite predictions," in the sense that predictions are always in principle available. It is just that ever increasing degrees of sophistication would be necessary to extract those predictions. The prediction process would never become routine.
 
Clearly, this would not be a particularly desirable state of affairs. But it would seem to be merely an inconvenience ---far from a disaster for physics.
 
The status of the other criteria for physical theories arose, not through general arguments, but rather through experience with specific examples. Whether physical theories will be required to possess some algorithm by which they can be implemented will presumably be settled by the same process.

\eject

\acknowledgements

Many colleagues gave helpful advice on a preliminary version of this paper. We are grateful in particular to B. DeWitt, J. Friedman, B. Horne, G. Horowitz, W. Kohn, A. Komar, G. Kreisel, D. Scalapino, R. Sorkin, M. Srednicki, and S. Wolfram. This work was supported in part by the NSF.

\end{document}